\documentclass[letter]{aa}  

\usepackage{graphicx}
\usepackage[varg]{txfonts}
\usepackage{hyperref}

\newcommand{\R}{\mathbb{R}}

\title{Observable properties of strong gravitational lenses}

\author{Nicolas Tessore}

\institute{%
Jodrell Bank Centre for Astrophysics, University of Manchester,
Alan Turing Building, Oxford Road, Manchester, M13 9PL, UK \\
\email{nicolas.tessore@manchester.ac.uk}
}

\date{Received 24 October 2016 / Accepted 15 November 2016}

\abstract{%
It is shown which properties of a strong gravitational lens can in principle be recovered from observations of multiple extended images when no assumptions are made about the deflector or sources.
The mapping between individual multiple images is identified as the carrier of information about the gravitational lens and it is shown how this information can be extracted from a hypothetical observation.
The derivatives of the image map contain information about convergence ratios and reduced shears over the regions of the multiple images.
For two observed images, it is not possible to reconstruct the convergence ratio and shear at the same time.
For three observed images, it is possible to recover the convergence ratios and reduced shears identically.
For four or more observed images, the system of constraints is overdetermined, but the same quantities can theoretically be recovered.
}


\begin{document} 

\maketitle

\section{Introduction}

In a recent article, \cite{2016A&A...590A..34W} have shown that derivatives and ratios of derivatives of the lensing potential near critical curves can be constrained directly from observations of strong gravitational lenses without the reconstruction of a lens model.
This raises the question of whether or not other lensing quantities can, at least in principle, be constrained from observations alone.
In this letter, an answer is given by explicit construction of the observable properties of strong gravitational lenses.
Observables are quantities which can be recovered from observations without the assumption of a specific lens model or reconstruction method, and are necessarily invariant under the mass sheet transform \citep[MST,][for a summary see \citealp{2006glsw.conf.....M}]{1985ApJ...289L...1F,1988ApJ...327..693G,1995A&A...294..411S}.
The approach presented is a general recipe for the extraction of model-independent information from observations of strong lenses with multiple extended images, which is applied here to recover the convergence and shear.

The structure of the letter is as follows.
Section~\ref{sec:image-map} introduces the mapping between multiple images as the essential holder of the observation's information about a strong gravitational lens.
Section~\ref{sec:convergence-shear} shows how observables can be reconstructed from the observed image map.
Section~\ref{sec:multiple-images} generalises the result to more than two observed images and treats the case in which some lens properties can be recovered identically.
Section~\ref{sec:discussion} closes with a brief discussion of the idea presented here and a look towards its possible application with future observations.

\section{The image map}
\label{sec:image-map}

\begin{figure}%
\centering%
\includegraphics[width=\columnwidth]{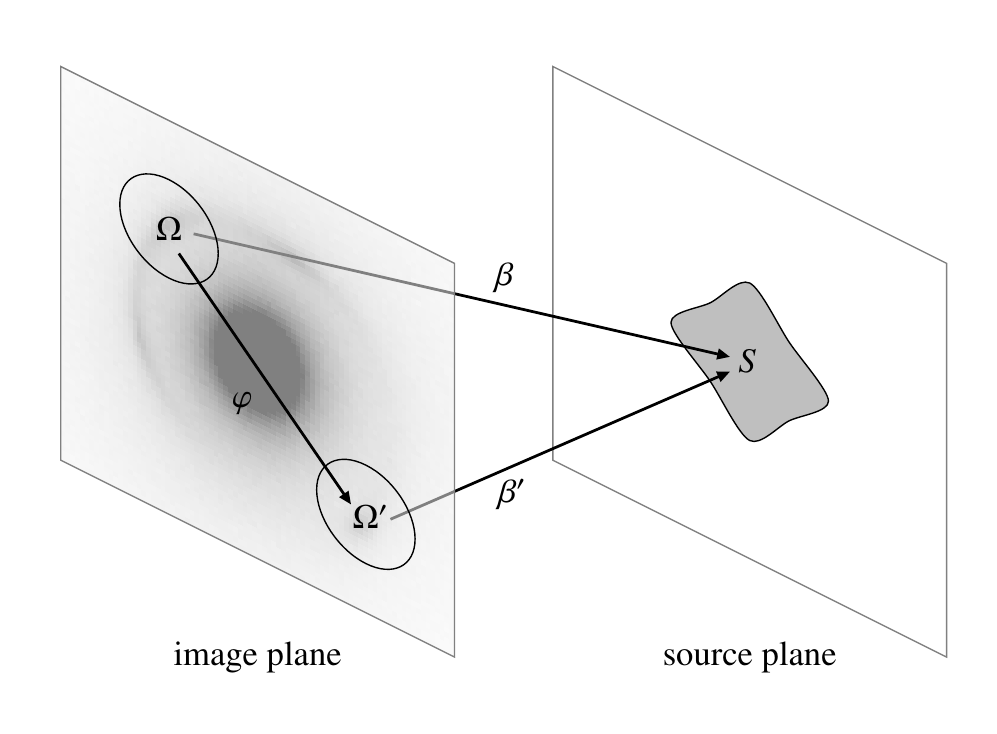}%
\caption{%
An observation shows multiple images $\Omega$ and $\Omega'$ of the source plane region $S$.
The image map $\varphi \colon \Omega \to \Omega'$ connects those points $\vec\theta \in \Omega$ and $\vec\theta' \in \Omega'$ that show the same point in the source plane, and can be used to relate the source position mappings $\beta'$ and $\beta = \beta' \circ \varphi$.
}%
\label{fig:image-map}%
\end{figure}

When no assumptions are made about the mass distribution of the deflector or about the surface brightness distribution of the lensed sources, any inference about a single strong gravitational lens must be based on the observation of multiple images of the background objects.
Consider a hypothetical ideal observation of two images $\Omega$ and $\Omega'$ of the same source plane region (Fig.~\ref{fig:image-map}).
The most information that can be extracted, even in principle, from the two images $\Omega$ and $\Omega'$ is a map $\varphi$ which connects those points $\vec\theta \in \Omega$ and $\vec\theta' \in \Omega'$ that show the same point in the source plane.
This image map must therefore contain all the properties of the gravitational lens that can be reconstructed, and observable properties of a strong gravitational lens are precisely those which can be recovered from the image map.

To formally establish the relationship between image map~$\varphi$ and the gravitational lens, let $\Omega, \Omega' \subset \R^2$ be two regions in the image plane, each of which shows a multiple image of the same region $S \subset \R^2$ in the source plane.
Considering the lens equation separately for~$\Omega$ and~$\Omega'$, let $\beta$ and $\beta'$ be the individual source position mappings
\begin{gather}
    \beta \colon \Omega \to S, \;
    \vec\theta \mapsto \vec\beta(\vec\theta) = \vec\theta - \vec\alpha(\vec\theta) \;, \label{eq:spm} \\
    \beta' \colon \Omega' \to S, \;
    \vec\theta' \mapsto \vec\beta'(\vec\theta') = \vec\theta' - \vec\alpha(\vec\theta') \;, \label{eq:spm1}
\end{gather}
where the regions $\Omega$ and $\Omega'$ are suitably restricted to contain only the multiple images (hence $\beta(\Omega) = \beta'(\Omega') = S$), and $\alpha$ is the deflection angle of the gravitational lens.
The image map~$\varphi$ is a mapping between the multiple images $\Omega$ and $\Omega'$,
\begin{equation}\label{eq:phi}
    \varphi \colon \Omega \to \Omega', \;
    \vec\theta \mapsto \vec\theta' \;,
\end{equation}
where $\vec\theta$ and $\vec\theta' = \vec\varphi(\vec\theta)$ map to the same point $\vec\beta(\vec\theta) = \vec\beta'(\vec\theta')$ in the source plane.
It follows that there is a relation between image map~$\varphi$ and the source position mappings,%
\begin{equation}\label{eq:diag}%
\arraycolsep=0pt%
\begin{array}{ccc}%
\Omega         & \xrightarrow{~~~\varphi~~} & \Omega'     \\
~~_{\beta}     &    \searrow~~~~\swarrow    & _{\beta'}~~ \\
               &              S             &
\end{array} \;,
\end{equation}
where~$\beta$ and~$\beta'$ are determined by the gravitational lens.
The equivalence between the source position mappings,
\begin{equation}\label{eq:betarel}
    \beta = \beta' \circ \varphi \;,
\end{equation}
can now be used to infer the properties of the lens from the image map by comparing the left and right-hand sides under various operations.

\section{Convergence and shear}
\label{sec:convergence-shear}

Equation~\eqref{eq:betarel} can be expanded to first order on both sides, which yields a relation,
\begin{equation}\label{eq:jacmat}
	\tens A(\vec\theta)
	= \tens A(\vec\varphi(\vec\theta)) \, \tens J_{\varphi}(\vec\theta) \;,
\end{equation}
between the magnification matrix~$\tens A$ evaluated at $\vec\theta$ and $\vec\theta' = \vec\varphi(\vec\theta)$, respectively, and the Jacobian matrix~$\tens J_\varphi$ of the image map.
It follows that $\tens J_\varphi$ is equal to the relative magnification matrix,
\begin{equation}
	\tens T(\vec\theta)
	= \tens A^{-1}(\vec\varphi(\vec\theta)) \, \tens A(\vec\theta) \;,
\end{equation}
between the multiple images~$\Omega$ and $\Omega'$.
The matrix~$\tens T$ is the only model-independent information that can be extracted from the observation at first order.

Writing the magnification matrix~$\tens A$ in terms of convergence~$\kappa$ and the two components~$g_1, g_2$ of the reduced shear, the matrix equation $\tens J_\varphi = \tens T$ can be split into partial differential equations,
\begin{gather}
	\frac{\partial\varphi_1}{\partial\theta_1}
	= \frac{1 - \kappa}{1 - \kappa'} \, \frac{(1 - g_1^{})(1 + g_1') - g_2^{} \, g_2'}{1 - g_1'^2 - g_2'^2} \;, \label{eq:d1} \\
	\frac{\partial\varphi_1}{\partial\theta_2}
	= \frac{1 - \kappa}{1 - \kappa'} \, \frac{(1 + g_1^{}) \, g_2' - (1 + g_1') \, g_2^{}}{1 - g_1'^2 - g_2'^2} \;, \label{eq:d2} \\
	\frac{\partial\varphi_2}{\partial\theta_1}
	= \frac{1 - \kappa}{1 - \kappa'} \, \frac{(1 - g_1^{}) \, g_2' - (1 - g_1') \, g_2^{}}{1 - g_1'^2 - g_2'^2} \;, \label{eq:d3} \\
	\frac{\partial\varphi_2}{\partial\theta_2}
	= \frac{1 - \kappa}{1 - \kappa'} \, \frac{(1 + g_1^{})(1 - g_1') - g_2^{} \, g_2'}{1 - g_1'^2 - g_2'^2} \label{eq:d4} \;,
\end{gather}
that generate the image map~$\varphi$, where the unprimed and primed values are taken at $\vec\theta \in \Omega$ and $\vec\theta' = \vec\varphi(\vec\theta) \in \Omega'$, respectively.

The observable lens properties from two images are then the quantities which can be constructed from combinations of the image map derivatives~\eqref{eq:d1} to~\eqref{eq:d4}.
For example, the Jacobian determinant of the image map is the local magnification ratio,
\begin{equation}\label{eq:jac}
    \det \tens J_\varphi
    = \frac{(1 - \kappa)^2}{(1 - \kappa')^2} \, \frac{1 - g_1^2 - g_2^2}{1 - g_1'^2 - g_2'^2}
    = \frac{\mu(\vec\theta')}{\mu(\vec\theta)} \;,
\end{equation}
at corresponding points $\vec\theta \in \Omega$ and $\vec\theta' = \vec\varphi(\vec\theta) \in \Omega'$.
This result is well-known and could be obtained more directly by taking the determinant on both sides of equation~\eqref{eq:jacmat}.

Unfortunately, no combination of derivatives~\eqref{eq:d1} to~\eqref{eq:d4} will be able to recover the convergences~$\kappa$ and~$\kappa'$ directly, since the equations only contain the convergence ratio,
\begin{equation}\label{eq:f}
	f = \frac{1 - \kappa}{1 - \kappa'} \;,
\end{equation}
and neither $\kappa$ nor $\kappa'$ in isolation.
This also follows from the MST; the convergence~$\kappa$ is not an invariant and therefore cannot be observable.
The convergence ratio~$f$, on the other hand, is an invariant, which makes it a prime candidate for an observable.
However, equations~\eqref{eq:d1} to~\eqref{eq:d4} still provide only four constraints for the five remaining unknowns $f, g_1^{}, g_2^{}, g_1', g_2'$.
The image map between two multiple images on its own is hence not enough to recover the convergence ratio and shear at the same time.

While it is tempting to augment the system of constraints by considering the magnification ratio~$\mu(\vec\theta')/\mu(\vec\theta)$ as observable, this would be incorrect.
For the assumed ideal observation, the magnification ratio must be constructed explicitly by mapping areas between the multiple images.
This is the same information contained in the image map and it cannot be used twice.
The correct way to construct more informative observables is to use more than two multiple images in the reconstruction.

\section{Multiple images}
\label{sec:multiple-images}

Consider now $n > 2$ multiple images $\Omega_0, \Omega_1, \dots, \Omega_{n-1}$ of the same source plane region~$S$.
Keeping~$\Omega_0$ fixed, it is possible to construct $n-1$ individual image maps,
\begin{equation}
	\varphi_i \colon \Omega_0 \to \Omega_i \;, \quad
	i = 1, \dots, n-1 \;,
\end{equation}
which contain all the information about the multiple images, as any further map $\varphi_{ij} \colon \Omega_i \to \Omega_j$ is the composition $\varphi_{ij} = \varphi_j \circ \varphi_i^{-1}$.
Every pair of images $(\Omega_0, \Omega_i)$ obeys a relation similar to~\eqref{eq:diag},
\begin{equation}\label{eq:diag_n}%
\arraycolsep=0pt%
\begin{array}{ccc}%
\Omega_0     & \xrightarrow{~~~\varphi_i~~} & \Omega_i     \\
~~_{\beta_0} &     \searrow~~~~\swarrow     & _{\beta_i}~~ \\
             &               S              &
\end{array} \;, \quad
i = 1, \dots, n-1 \;,
\end{equation}
between its image map~$\varphi_i$ and suitably restricted source position mappings~$\beta_0 \colon \Omega_0 \to S$ and $\beta_i \colon \Omega_i \to S$.
In particular, there are now $n-1$ individual equations that relate $\beta_0$ and $\beta_i$ in the same manner as equation~\eqref{eq:betarel} did in the case of two images,
\begin{equation}\label{eq:betarel_n}
    \beta_0 = \beta_i \circ \varphi_i \;, \quad
    i = 1, \dots, n-1 \;.
\end{equation}
Hence there is a set of equations~\eqref{eq:d1} to~\eqref{eq:d4} for each multiple image $i = 1, \dots, n-1$,
\begin{gather}
	\frac{\partial\varphi_{i,1}}{\partial\theta_1}
	= \frac{1 - \kappa}{1 - \kappa_i} \, \frac{(1 - g_{0,1})(1 + g_{i,1}) - g_{0,2} \, g_{i,2}}{1 - g_{i,1}^2 - g_{i,2}^2} \;, \label{eq:d1_n} \\
	\frac{\partial\varphi_{i,1}}{\partial\theta_2}
	= \frac{1 - \kappa}{1 - \kappa_i} \, \frac{(1 + g_{0,1}) \, g_{i,2} - (1 + g_{i,1}) \, g_{0,2}}{1 - g_{i,1}^2 - g_{i,2}^2} \;, \label{eq:d2_n} \\
	\frac{\partial\varphi_{i,2}}{\partial\theta_1}
	= \frac{1 - \kappa}{1 - \kappa_i} \, \frac{(1 - g_{0,1}) \, g_{i,2} - (1 - g_{i,1}) \, g_{0,2}}{1 - g_{i,1}^2 - g_{i,2}^2} \;, \label{eq:d3_n} \\
	\frac{\partial\varphi_{i,2}}{\partial\theta_2}
	= \frac{1 - \kappa}{1 - \kappa_i} \, \frac{(1 + g_{0,1})(1 - g_{i,1}) - g_{0,2} \, g_{i,2}}{1 - g_{i,1}^2 - g_{i,2}^2} \label{eq:d4_n} \;,
\end{gather}
where the convergence~$\kappa_i$ and shear~$g_{i,1}, g_{i,2}$ for each image are evaluated at the point $\vec\varphi_i(\vec\theta)$, respectively.

The goal is now to solve the system of equations~\eqref{eq:d1_n} to~\eqref{eq:d4_n} for the lens observables.
The total number of constraints is $N_\text{con} = 4n - 4$.
For the reasons given above, it is necessary to introduce the convergence ratio for each image,
\begin{equation}\label{eq:f_n}
	f_i = \frac{1 - \kappa_0}{1 - \kappa_i} \;, \quad
	i = 1, \dots, n-1 \;,
\end{equation}
since the $n$ convergences $\kappa_0, \kappa_1, \dots$ are not directly observable.
The observables are then $g_{0,1}, g_{0,2}, f_1, g_{1,1}, g_{1,2}, \dots$, and the total number of unknowns is $N_\text{obs} = 3n - 1$.
The system of equations therefore admits three distinct cases.

For $n = 2$ multiple images, $N_\text{con} < N_\text{obs}$ and the system is underdetermined.
This is seen in Section~\ref{sec:convergence-shear}.

For $n = 3$ multiple images, $N_\text{con} = N_\text{obs}$ and the system is solvable.
The lens properties can hence be recovered identically from the image maps by solving equations~\eqref{eq:d1_n} to~\eqref{eq:d4_n} for the observables.
Taking the combinations
\begin{gather}
	a_i = \frac{\partial\varphi_{i,1}}{\partial\theta_1} - \frac{\partial\varphi_{i,2}}{\partial\theta_2} \;, \\
	b_i = \frac{\partial\varphi_{i,2}}{\partial\theta_1} + \frac{\partial\varphi_{i,1}}{\partial\theta_2} \;, \\
	c_i = \frac{\partial\varphi_{i,2}}{\partial\theta_1} - \frac{\partial\varphi_{i,1}}{\partial\theta_2} \;, \\
	d_i = \frac{\partial\varphi_{i,1}}{\partial\theta_1} + \frac{\partial\varphi_{i,2}}{\partial\theta_2} \;,
\end{gather}
of the derivatives of the two image maps ($i = 1, 2$), the reduced shear $g_0$ over multiple image $\Omega_0$ is found to be
\begin{gather}
	g_{0,1} = \frac{a_1 c_2 - a_2 c_1}{b_1 a_2 - b_2 a_1} \;, \\
	g_{0,2} = \frac{b_1 c_2 - b_2 c_1}{b_1 a_2 - b_2 a_1} \;.
\end{gather}
For the additional multiple images $\Omega_i$, $i = 1,2$, the convergence ratio~$f_i$ and reduced shear $g_i$ are most easily expressed in terms of the results for $g_{0,1}$ and $g_{0,2}$ as
\begin{gather}
	f_i~~\, = \frac{2 J_i}{a_i \, g_{0,1} + b_i \, g_{0,2} + d_i} \;, \\
	g_{i,1} = \frac{d_i \, g_{0,1} - c_i \, g_{0,2} + a_i}{a_i \, g_{0,1} + b_i \, g_{0,2} + d_i} \;, \\
	g_{i,2} = \frac{c_i \, g_{0,1} + d_i \, g_{0,2} + b_i}{a_i \, g_{0,1} + b_i \, g_{0,2} + d_i} \;,
\end{gather}
where $J_i = (c_i^2 + d_i^2 - a_i^2 - b_i^2)/4$ is the Jacobian determinant of image map~$\varphi_i$.

For $n > 3$ multiple images, $N_\text{con} > N_\text{obs}$ and the system is overdetermined.
In this case, all lens properties can, in principle, still be recovered from observations.
The additional constraints are a consistency check of the reconstruction, and could be used to test image maps constructed from observations.

\section{Conclusions}
\label{sec:discussion}

It has been known since the discovery of the MST that many properties of a gravitational lens are not observable.
The results of Section~\ref{sec:multiple-images} now explicitly show properties that can, at least in principle, be observed from multiple extended images.
This is of practical interest:
knowing which properties of a lens system are effectively constrained by the observation is necessary to present significant results.
Reconstructions should compare maps of the convergence ratios $f_1, f_2, \dots$ and reduced shears $g_0, g_1, \dots$, as these are the effectively observable properties of strong lenses.

As it stands, the image map~$\varphi$ is a theoretical construct that is introduced here as a tool to find properties of gravitational lenses that can be recovered from multiple observed images.
However, it is possible that future telescopes have the necessary resolution to make a point-by-point mapping between multiple images of the largest strong lenses feasible, at least to some degree.
In this case, the technique outlined here could lead to a truly model-free reconstruction method for strong lenses that uses no more information than what is provided by observations.

\begin{acknowledgements}
The author would like to thank M.~Bartelmann, S.~Bridle and R.~B.~Metcalf for their comments and discussions.
The anonymous referee helpfully suggested changing the derivation from using complex derivatives to the better-known real-valued expressions, which simplified the discussion.
The author further acknowledges support from the European Research Council in the form of a Consolidator Grant with number 681431.
\end{acknowledgements}

\bibliographystyle{aa}
\bibliography{nt1601}

\end{document}